# Resonant Structures in the Low-Energy Electron Continuum for Single Ionization of Atoms in the Tunneling Regime

A. Rudenko, K. Zrost, C.D. Schröter, V.L.B. de Jesus, B. Feuerstein,

R. Moshammer, J. Ullrich

Max-Planck-Institut für Kernphysik, Saupfercheckweg 1, D-67119 Heidelberg, Germany

We present results of high-resolution experiments on single ionization of He, Ne and Ar by ultra-short (25 fs, 6 fs) 795 nm laser pulses at intensities $0.15 - 2.0 \cdot 10^{15}$ W/cm$^2$. We show that the ATI-like pattern can survive deep in the tunneling regime and that the atomic structure plays an important role in the formation of the low-energy photoelectron spectra even at high intensities. The absence of ponderomotive shifts, the splitting of the peaks and their degeneration for few-cycle pulses indicate that the observed structures originate from a resonant process.

Single electron emission from atoms exposed to intense short-pulse laser fields is one of the most fundamental and well-studied reaction channels in non-linear strong-field physics [1]. One of the basic questions, however, which still disturbs a comprehensive understanding of the ionization dynamics, concerns the intensity regime that might be characterised by the transition from a "photon" to a "field" based view of the problem. Usually, ionization is considered as either a "multiphoton" or "tunnelling" process depending on the value of the so-called Keldysh parameter $\gamma = \sqrt{I_p/2U_p}$ (where $I_p$ is the ionization potential and $U_p = I/4\omega^2$ the ponderomotive potential. I is the light intensity and $\omega$ its frequency; atomic units are used throughout). For $\gamma > 1$ numerous experiments revealed the multiphoton nature of ionization reflected in a rich structure in the photoelectron spectra. Various field induced processes, such as resonant and non-resonant above-threshold ionization (ATI), ponderomotive effects, channel-opening and closing etc. were studied, and good agreement between theory and experiment has been achieved [2-7].

For the other extreme case of very low $\gamma$ only few experimental results have been reported. Early experiments performed with low-frequency radiation [8] showed a good agreement with a quasistatic tunnelling model [9], which assumes that the electron first tunnels into the continuum and then moves classically driven by the oscillating laser field. Later, experiments at optical frequencies observed a significant fraction of electrons with energies higher than the maximal drift energy of 2 $U_p$ that an electron can gain in the field, forming a plateau-like structure in the high-energy part of the electron spectra [10,11]. Since then, efforts strongly concentrated to explore this feature, which has been explained within the tunnelling picture by considering the return of the electron to the ion core and subsequent elastic rescattering [11].

Recently, high-resolution experiments [12,13] provided evidences for the resonant origin of detailed narrow structures within the plateau, in agreement with a series of numerical simulations [14,15]. A comprehensive analysis of the problem performed in [15] revealed a close relation between resonant-like behaviour within the plateau and the existence of electron trajectories which lead to multiple recollisions with the parent ion [16]. The significance of multiphoton resonances for "super-ponderomotive" photoelectrons in the tunnelling regime was predicted in [17] for the case of He, but was never confirmed experimentally.

The lowest value of the Keldysh parameter for which some regular ATI structure was ever observed is, to the best of our knowledge, $\gamma \sim 0.7$ [6]. Experiments performed at lower $\gamma$ either revealed a smooth decrease of the photoelectron spectra, with changes of the slope in the plateau region [8,11], or observed in addition some hardly resolved structure below 20 eV [6,18]. Very recently precise measurements of ultra-low energy electron emission found a clear double-peak structure around zero [19] in the electron momentum distribution parallel to the laser polarization for Ne at the intensity $1 \cdot 10^{15}$ W/cm$^2$ ($\gamma = 0.42$). This minimum was predicted by recent semiclassical calculations [20] and was related to the Coulomb interaction of the emitted electron with the parent ion.

In this Letter we report on the first experimental observation of well resolved pronounced and regular pattern in the electron emission characteristics for single ionization of He, Ne and Ar at intensities of up to $1.5 \cdot 10^{15}$ W/cm$^2$ ($\gamma < 0.5$). We show that for 25 fs laser pulses the low-energy part of the photoelectron spectra consists of a series of wide peaks separated by the photon energy, which exhibit features (absence of ponderomotive shifts, narrow substructures resolved within different peaks) typical for resonantly-enhanced ionization. These features are considerably washed out for 6-7 fs pulses. For ultra-low energy electrons we observe clear signatures of the atomic structure, while for higher energies the spectra are similar for different atomic species.

The experiments were performed using a newly designed "reaction microscope" [21] with further improved momentum resolution along the laser polarization direction ($\Delta P_\parallel <$ 0.02 a.u.). We used linearly polarized radiation of a Kerr-lens mode locked Ti:sapphire laser at 795 nm wavelength amplified to pulse energies of up to 350 µJ at 3 kHz. The width of the amplified pulses was 25 fs. To generate few-cycle pulses they are spectrally broadened in a gas-filled hollow fiber and then compressed to 6-7 fs by chirped mirrors and a prism compressor. The laser beam was focused to a spot size of ~ 7 µm on the collimated supersonic gas jet in the ultra-high vacuum chamber ($2 \cdot 10^{-11}$ mbar). Fluctuations of the laser intensity from pulse to pulse were monitored during the experiment and did not exceed 5%. Absolute calibration of the peak intensity was performed using a clear kink in the measured photoelectron momentum distribution, which corresponds to the maximum drift momentum of $2\sqrt{U_p}$ that electrons can gain from the laser field (inset of Fig.1, for details see [22]).

Ions and electrons were guided to two position-sensitive channel plate detectors by weak electric (1V/cm) and magnetic (5G) fields applied along the laser polarization axis. From the time of flight and position the full momentum vectors of coincident ions and electrons were calculated. Neglecting the small momentum of the absorbed photons, momentum conservation was ensured for each single event. This allows one to exclude possible influences of space-charge effects or ponderomotive acceleration in the spatially inhomogeneous laser field since both would act differently on electrons and ions, and ensures that absolutely no contributions from higher charge states or impurities enter the spectra.

Fig. 1 displays momentum distributions of singly charged ions along the laser polarization axis for He, Ne and Ar at different intensities ($\gamma$ varies from 1.1 to 0.29) for pulse lengths of 25 fs (Fig. 1a-c) and 6 fs (Fig. 1d), respectively. The data are integrated over the transverse momenta. For all three targets the spectra at the lowest intensity manifest a clear set of peaks that are broadened as the intensity grows, emerging for the highest intensities into a shoulder-like structure at the positions of the first two peaks. Several surprising features can be observed. First, at least remnants of the lowest order peaks can be clearly followed up to the highest intensities, deep in the tunneling regime. Second, all peak positions are independent of the intensity. Third, clear remnants of the peaks can be observed even for the 6 fs pulses, although they are considerably smoothed.

Apart from the overall similarity of the observed pattern for all the targets there is one striking difference, the appearance of a pronounced minimum at zero longitudinal momentum for He and Ne, whereas there is a clear maximum for Ar. Within semiclassical calculations [20], the dip at zero momentum was attributed to the interaction of electrons with their parent ion through many recollisions at very low energy. Accordingly, the shape of the scattering potential should influence the emission of ultra-low energy electrons and therefore they might be the only ones that remain sensitive to the target structure at very high intensities.

In order to clarify the origin of the observed structures, we next examine the electron energy spectra. As it is shown in Fig. 2 for the case of Ne, electron energy distributions also exhibit a series of peaks, which are separated by the photon energy and broadened with increasing intensity. Thus, we do observe the evolution of the ATI-like structure and again, as in the momentum representation, the peaks do not exhibit intensity-dependent shifts (see vertical dashed lines in Fig. 1-2). Zooming into the details by fully exploiting the high resolution of our electron imaging system along the longitudinal direction and plotting the longitudinal part of the electron kinetic energy only, we find a pronounced fine structure with at least two sub-peaks repeated within different ATI orders at different intensities (Fig. 2b).

The splitting of the ATI-like peaks together with the absence of any detectable pondermotive shifts might lead to assume them to be of resonant nature. Non-resonant ATI peaks are known to shift towards lower energies as the intensity grows due to the field–induced increase of the ionization potential, which is approximately equal to the ponderomotive energy $U_p$. For $\lambda = 795$ nm at a laser peak intensity of 1.0 PW/cm$^2$ $U_p$ is as high as 60 eV. Thus, averaging over different intensities (spatio-temporal pulse distribution and intensity fluctuations) on a level of less than 3% will result in variations of $U_p$ exceeding the photon energy, so that any non-resonant ATI peaks will definitely be smeared out.

The situation is different for the case of resonantly-enhanced ionization, which occurs through an excited atomic level shifted into multiphoton resonance with the ground state. In early studies two scenarios resulting in the intensity-independent ATI peak positions were suggested. First one [4] assumes that ionization from an excited state occurs before the intensity has considerably changed, so that it always proceeds at the intensity required to fulfil the resonance condition and accordingly, at some certain value of the ponderomotive shift. If the peak intensity is higher than the resonant value, the resonance condition can be fulfilled somewhere in the laser focus. The second scenario [5] suggests that the excited state survives longer in the laser pulse and is ionized later by one- or multi-photon absorption. If the excited state is a high-lying Rydberg state, it is shifted upwards almost as much as the continuum level, again giving rise to intensity-independent peak positions.

Of course, such considerations can hardly be directly applied at high intensities, where any intermediate state must be completely deformed by the laser field. However, numerical simulations [14,15,17] have shown that some excited states may remain surprisingly stable or even be induced by the laser field (light induced states). The part of the wave function that tunnels out of the atom might effectively populate these extended quasi-bound states producing resonant-like structures in the photoelectron spectra [13,17].

Contributions from any resonant-like mechanisms should be sensitive to the length of the laser pulse. For pulse durations of less than 10 fs all Rydberg states have Kepler orbits considerably longer, which can result in a suppression of resonantly-enhanced ionization [23]. Besides that, if the lifetime of the excited state is comparable to the pulse length, their mutual interrelation starts to play a role. In our data, comparing results obtained with 23 fs and 6-7 fs pulses for Ne (Figures 1b and 1d), one can see that with shorter pulses the ATI peaks are smoothed out, and are only distinguishable for the lowest intensity.

Even more information about the ionization dynamics can be obtained by consideration of the angular-resolved data. Figure 3 presents density plots of 2-dimensional electron momentum distributions for different atomic species and pulse lengths. In this representation ATI peaks manifest themselves as ring-like structures. In striking contrast to the common belief, the electron emission characteristics show a surprisingly rich structure, even in the tunneling regime. The spectra clearly contain target-specific features in the low-energy region; however, for higher longitudinal momenta the spectra for different targets exhibit similar pattern with pronounced maxima and minima in the angular distribution. All these features are noticeably washed out for short 6-7 fs pulses (compare Fig. 3c and 3d).

As of now, the origin of the structures observed in the two-dimensional momentum distributions in terms of simple pictures remains completely unclear. The angular distribution of any resonant–enhanced contribution should reflect the symmetry of the state through which ionization has occurred [7]. This can result in an atomic species dependence of the data, in particular, in different behaviour in the ultra-low energy region, as observed in Fig.3. Taking into account low peak-to-valley contrast of the spectra, we assume that the resonantly enhanced part is embedded into a strong nonresonant contribution, and thus, superposition of both should define the angular distribution of the emitted electrons.

In order to perform a detailed comparison between theory and experiment one should not only properly describe the atomic structure modified by the laser field, but also account for contributions from resonant and nonresonant intensities realized in different parts of the spatio-temporal distribution of the pulse. This way single ionization of Ar at intensities up to $10^{14}$ W/cm$^2$ was perfectly described within a single active electron model by solving numerically the time-dependent Schrödinger equation [7]. The extension of this approach to higher intensities will, of course, require larger computational efforts. However, our results show that truly quantum dynamical effects definitely play an important role even at high intensities, where the electron motion after the tunnelling is often viewed to proceed classically. If one considers tunnelling ionization as a periodic ejection of electron wave packets at the maxima of the oscillating laser field [24], quantum interferences between the coherent wave packets will lead to the well known ATI structure in the electron spectrum. For shorter pulses, only a few maxima contribute making the peak structure less pronounced.

In conclusion, we have studied the evolution of ATI-like structures in electron spectra over a broad range of Keldysh parameters ($1.1 > \gamma > 0.29$). We have shown that low-order ATI peaks with some target-dependent substructure can be resolved deep in the tunnelling regime. The peaks do not exhibit ponderomotive shifts and are less pronounced for ionization by few-cycle laser pulses. These are typical features for resonant-enhanced ionization, well-known for the case of much lower intensities. The importance of resonant enhancement and its connection to recollision phenomena was recently demonstrated for the plateau in the photoelectron spectra [12-17]. Taking into account latest results showing that recollision effects also manifest themselves at low electron energies [19,20], we conclude that the structures presented in our data originate from a similar type of resonant process. However, a detailed description of the intermediate states involved, their relation to the field-free atomic levels and influence on the electron energy and angular distributions remains the subject of forthcoming analyses.

This work was supported by the Leibniz-Program of the Deutsche Forschungsgemeinschaft. We are grateful to A. Voitkiv and B. Najjari for fruitful discussions and to K. Dimitriou and J. Burgdörfer for making their results available prior to publication.

**Figure captions:**

Fig. 1: Longitudinal momentum distributions of singly charged ions for different intensities. The curves are stretched by arbitrary factors in the vertical direction for visual convenience. (a) He, 25 fs. (b) Ar, 25 fs. (c) Ne, 25 fs. (d) Ne, 6-7 fs. The inset in (d) shows in logarithmic scale the spectrum used for the absolute intensity calibration (Ne, 25 fs, 0.4 PW/cm$^2$). Vertical lines there indicate the maximum drift momentum $\pm 2\sqrt{U_p}$.

Fig. 2: Electron energy distributions for single ionization of Ne by 25 fs pulses at different intensities. The data are integrated over all emission angles. (a) Total kinetic energy. (b) Longitudinal component of the kinetic energy.

Fig. 3: Electron momentum distribution parallel ($P_{//} = P_z$) and perpendicular ($P_\perp = \sqrt{P_x^2 + P_y^2}$) to the laser polarization direction. (a) He, 25 fs, 0.6 PW/cm$^2$. (b) Ar, 25 fs, 0.5 PW/cm$^2$. (c) Ne, 25 fs, 0.6 PW/cm$^2$. (d) Ne, 6-7 fs, 0.5 PW/cm$^2$. Vertical cuts show regions where the spectrometer has no resolution in the transverse direction.

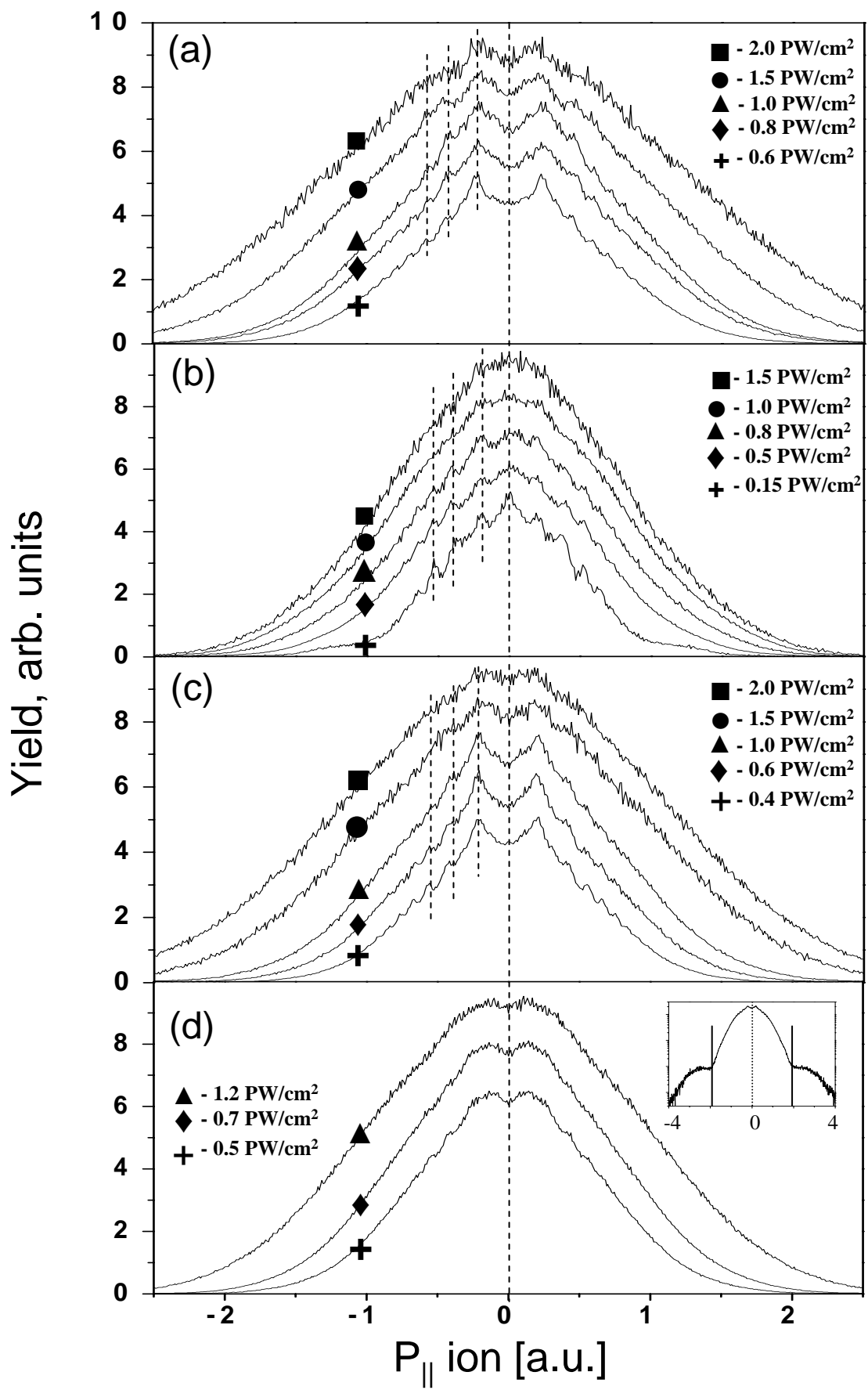

Figure 1

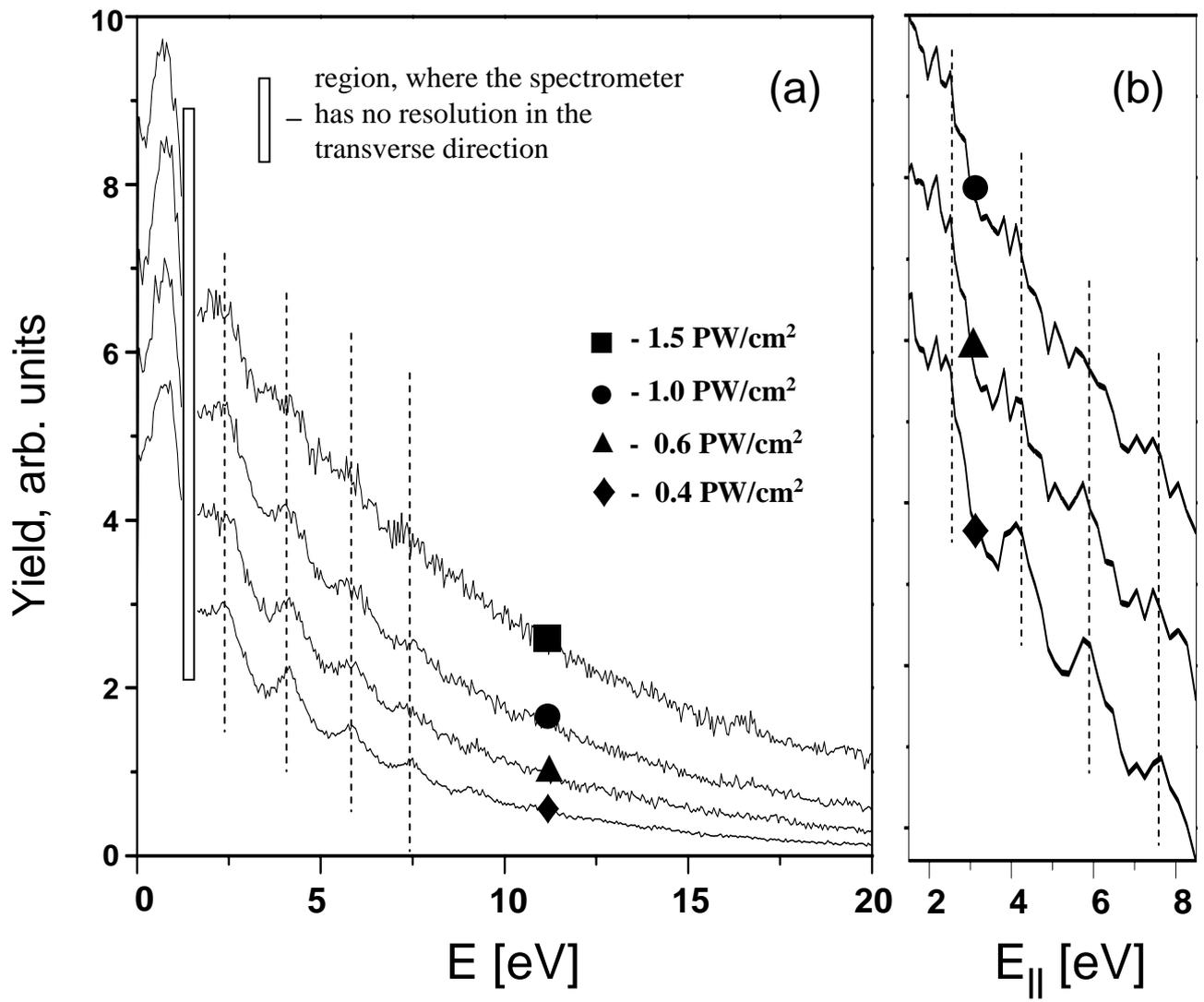

Figure 2

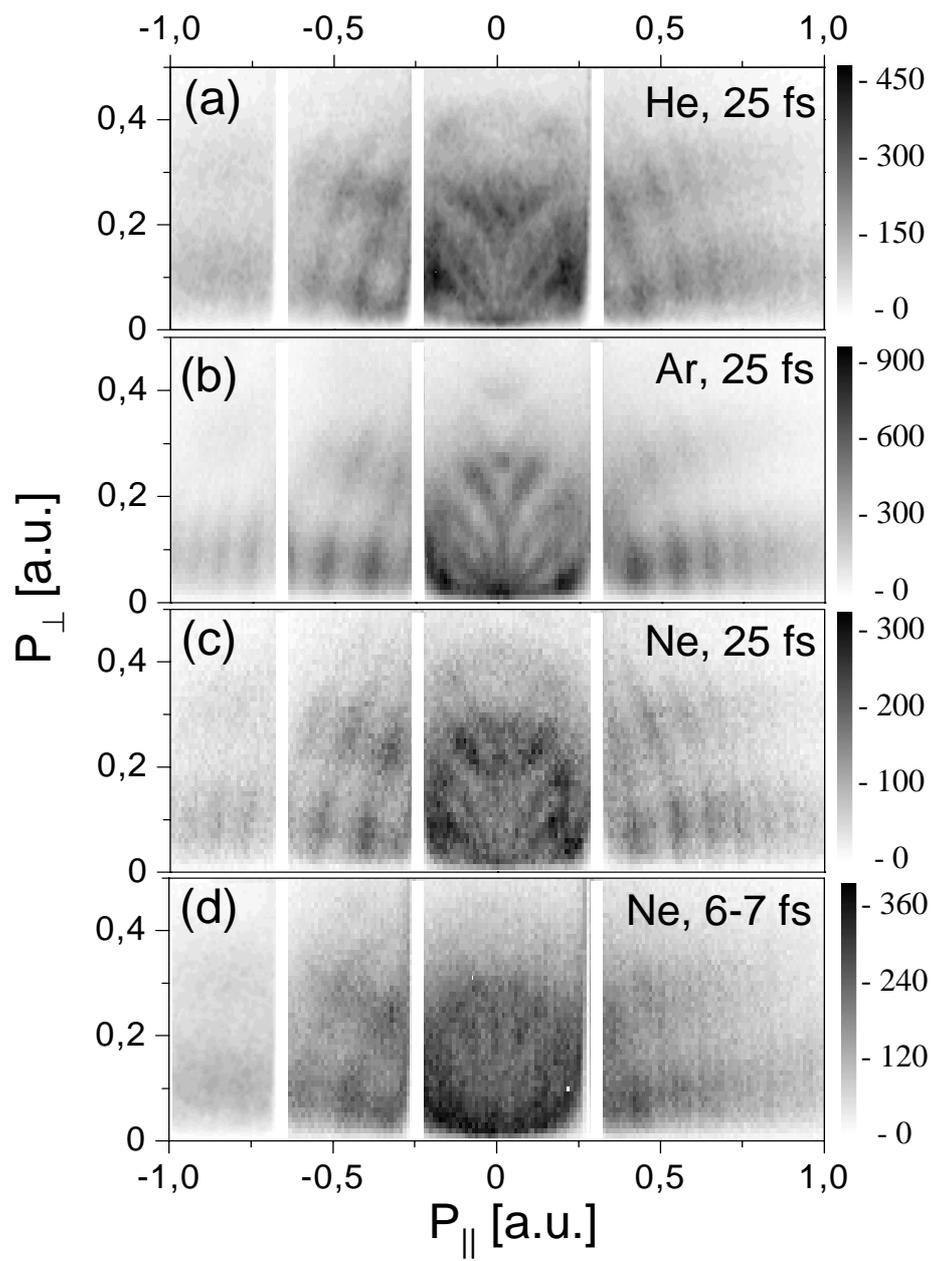

Figure 3